\documentclass[12pt,reqno]{amsart}
\usepackage{hyperref}
\usepackage{appendix}
\usepackage{graphicx}
\usepackage{amscd}
\usepackage{amsmath}
\usepackage{epsfig}
\usepackage{amsfonts}
\usepackage{amssymb}

\newtheorem{theorem}{Theorem}

\input{tcilatex}

\begin{document}
\title[\textquotedblleft Sommerfeld Puzzles\textquotedblright ]{The
\textquotedblleft Sommerfeld Puzzle\/\textquotedblright \\
and Its Extensions}
\author{Sergei K.~Suslov}

\begin{abstract}
The exact agreement between the Sommerfeld (1916) and Dirac (1928) results
for the energy levels of the relativistic hydrogen atom (the so-called
\textquotedblleft Sommerfeld puzzle\/\textquotedblright ) is analyzed and
extended. Werner Heisenberg called this coincidence a `miracle' but Erwin
Schr\"{o}dinger described it as a fortuitous computational accident.
\end{abstract}

\address{School of Mathematical and Statistical Sciences, Arizona State
University, P.~O. Box 871804, Tempe, AZ 85287-1804, U.~S.~A.}
\email{sks@asu.edu}
\maketitle

\noindent \textit{If you want to be a physicist, you must do three things -
first, study mathematics, second, study more mathematics, and third, do the
same. \smallskip }

\noindent \qquad \qquad \qquad \qquad \qquad \qquad\qquad \qquad \qquad
\qquad \qquad Arnold Sommerfeld

\section{Introduction}

One of the central problems in quantum mechanics is the spectrum of
hydrogenlike atoms. The fine-structure of hydrogen atom spectral lines was
discovered by Albert A.~Michelson in 1887: When his ether-wind experiments
have failed, he turned to spectroscopy and found that the leading $H_{\alpha
}$ line of the Balmer series comprises a doublet. (The electron was
discovered by J.~J.~Thomson in 1897 and Rutherford's model of
the atom appeared in 1911!)

In 1916 Arnold Sommerfeld \cite{Som1916}\ applied the quantization rules of
the `old' quantum theory to the relativistic hydrogen atom. Exact solution
was obtained by C.~G.~Darwin \cite{Darwin} and W.~Gordon \cite{Gordon}\ only
in 1928 after discovery of the Dirac equation \cite{DiracI}, \cite{DiracII},
\cite{DiracQM}, \cite{Kr:Lan:Sus16}: The new answer was precisely the `old'
Sommerfeld formula!

The puzzle had been discussed further in \cite{Biedenharn}, \cite{Gran},
\cite{Heisenberg}, \cite{Petrov}, and \cite{Vick}\ with different
interpretations from mathematics, physics, and philosophy.
Here, we approach this matter in a systematic and analytic fashion,
with some extensions to similar problems in quantum physics.

This work is dedicated to Professor Viktor V.~Dodonov on the occasion of his
seventy-fifth birthday (diamond jubilee).

\section{The original \textquotedblleft Sommerfeld Puzzle\/\textquotedblright%
}

\subsection{WKB approximation in Coulomb problems}

Phenomenological quantization rules of `old' quantum mechanics \cite{Som1916}%
, \cite{Wil1915} are derived in modern physics from the corresponding wave
equations in the so-called semiclassical approximation (WKB method) \cite%
{BerryMount}, \cite{Ghatetal}, \cite{Langer1931}, \cite{Langer1937}, \cite%
{NU}, \cite{Schiff}. (See also \cite{Kragh}, \cite{Mil2004}, \cite{SomAS},
and \cite{SomQM} for historical reviews.)

In this approximation, for a particle in the central field, one can use a
generic radial equation of the form\/:%
\begin{equation}
u^{\prime \prime }(x)+q(x)\,u=0\/,  \label{WKB1}
\end{equation}%
where $x^{2}\,q(x)$ is continuous together with its first and second
derivatives (see below) for $0\leq x\leq b<\infty \/.$ As is well-known, the
traditional semiclassical approximation cannot be used in a neighborhood of $%
x=0\/.$ However, the change of variables $x=e^{z}\/$, $u=e^{z/2}\,v(z)$
transforms the equation into a new form,%
\begin{equation}
v^{\prime \prime }(z)+q_{1}(z)\,v=0\/,  \label{WKB2}
\end{equation}%
where%
\begin{equation}
q_{1}(z)=-\dfrac{1}{4}+\left( x^{2}\,q(x)\right) _{x=e^{z}}\/  \label{WKB3}
\end{equation}%
(Langer's modification \cite{BerryMount}, \cite{Langer1937}, \cite{NU}). As $%
z\rightarrow -\infty $ (or $x\rightarrow 0$), the new function $q_{1}(z)$ is
changing slowly near the constant%
\begin{equation*}
-1/4+\lim_{x\rightarrow 0}x^{2}q(x)\/\qquad \text{and\quad }%
\lim_{z\rightarrow -\infty }q_{1}^{(k)}(z)=0\/ \quad (k=1,2).
\end{equation*}%
Hence the function $q_{1}(z)$ and its derivatives are changing slowly for
large negative $z\/.$

The WKB method can be applied to the new equation and, as a result, in the
original equation one should replace $q(x)$ with%
\begin{equation}
q(x)-1/(4x^{2})\/=p_{\text{effective}}^{2}(x)  \label{WKB4}
\end{equation}%
(see, for example, \cite{Langer1937} and \cite{NU} for more details).

For all Coulomb problems under consideration, one may utilize the following
generic integral, originally evaluated by Sommerfeld performing complex
integration \cite{SomAS}: If%
\begin{equation}
p(r)=\sqrt{-A+\dfrac{B}{r}-\dfrac{C}{r^{2}}}\;\qquad A\/,C>0\/.  \label{WKB5}
\end{equation}%
Then%
\begin{equation}
\int_{r_{1}}^{r_{2}}p(r)\,dr=\pi \left( \dfrac{B}{2\sqrt{A}}-\sqrt{C}\right)
\label{WKB6}
\end{equation}%
provided $p(r_{1})=p(r_{2})=0\/.$ (See \cite{Barleyetal2021} and appendix
for elementary evaluations of this integral.)

The Bohr--Sommerfeld quantization rule takes the form \cite{NU}, \cite%
{Schiff}:%
\begin{equation}
\int_{r_{1}}^{r_{2}}p(r)\,dr=\pi \left( n_{r}+\dfrac{1}{2}\right) \quad
\quad (n_{r}=0\/,1\/,2\/,\,\dots )  \label{WKB7}
\end{equation}%
and, for the energy levels, one gets the following equation:%
\begin{equation}
\dfrac{B}{2\sqrt{A}}-\sqrt{C}=n_{r}+\dfrac{1}{2}.  \label{WKB8}
\end{equation}%
It is worth noting that a variant of this relation was found by Sommerfeld
himself \cite{SomQM}, by complex integration, in his original attempt to
explain how the quantum rules of the `old'\ theory (but only in the cases of
the one-dimensional harmonic oscillator and the Kepler problem) are
connected with the wave mechanics of Schr\"{o}dinger.

\subsection{Exact solutions: Nikiforov-Uvarov approach}

Generalized equation of the hypergeometric type \cite{NU},%
\begin{equation}
u^{\prime \prime }+\frac{\widetilde{\tau }(x)}{\sigma (x)}u^{\prime }+\frac{%
\widetilde{\sigma }(x)}{\sigma ^{2}(x)}u=0  \label{NU1}
\end{equation}%
($\sigma ,$ $\widetilde{\sigma }$ are polynomials of degrees at most $2,%
\widetilde{\tau }$ is a polynomial at most first degree), by the substitution%
\begin{equation}
u=\varphi (x)y(x)  \label{NU2}
\end{equation}%
can be reduced to the form%
\begin{equation}
\sigma (x)y^{\prime \prime }+\tau (x)y^{\prime }+\lambda y=0  \label{NU3}
\end{equation}%
if:%
\begin{equation}
\frac{\varphi ^{\prime }}{\varphi }=\frac{\pi (x)}{\sigma (x)},\qquad \pi
(x)=\frac{1}{2}\left( \tau (x)-\widetilde{\tau }(x)\right)  \label{NU4}
\end{equation}%
(or, $\tau (x)=\widetilde{\tau }+2\pi ,$ for later),%
\begin{equation}
k=\lambda -\pi ^{\prime }(x)\qquad (\text{or, }\lambda =k+\pi ^{\prime }),
\label{NU5}
\end{equation}%
and%
\begin{equation}
\pi (x)=\frac{\sigma ^{\prime }-\widetilde{\tau }}{2}\pm \sqrt{\left( \frac{%
\sigma ^{\prime }-\widetilde{\tau }}{2}\right) ^{2}-\widetilde{\sigma }%
+k\sigma }  \label{NU6}
\end{equation}%
is a linear function. (Use the choice of constant $k$ to complete the square
under the radical symbol; see \cite{Barleyetal2021}, \cite{EEKS}, and \cite{NU}
for more details.)

In Nikiforov-Uvarov's method, the energy levels can be obtained from the
quantization rule:%
\begin{equation}
\lambda +n\tau ^{\prime }+\frac{1}{2}n\left( n-1\right) \sigma ^{\prime
\prime }=0\qquad (n=0,1,2,\ ...)  \label{NU7}
\end{equation}%
and the corresponding square-integrable solutions are classical orthogonal
polynomials, up to a factor. They can be found by the Rodrigues-type formula:%
\begin{equation}
y_{n}(x)=\dfrac{B_{n}}{\rho (x)}\left[ \sigma ^{n}(x)\rho (x)\right]
^{(n)},\qquad (\sigma \rho )^{\prime }=\tau \rho ,  \label{NU8}
\end{equation}%
where $B_{n}$ is a constant. Each infinite (countable) set of these
square-integrable solutions is complete \cite{NU}.

\subsection{The Sommerfeld-type potentials in the Nikiforov-Uvarov approach}

Let us choose%
\begin{eqnarray}
\sigma (x) &=&x,\qquad \widetilde{\tau }(x)=0,  \label{STP1} \\
\widetilde{\sigma }(x) &=&-ax^{2}+bx-c+\frac{1}{4}.  \notag
\end{eqnarray}%
Then%
\begin{equation}
\pi (x)=\frac{1}{2}\pm \sqrt{ax^{2}+(k-b)x+c}.  \label{STP2}
\end{equation}%
When $k=b\pm 2\sqrt{ac},$ one can complete the square and obtain%
\begin{equation}
\pi =\frac{1}{2}\pm \left( \sqrt{a}x\pm \sqrt{c}\right) ,\qquad \tau =2\pi .
\label{STP3}
\end{equation}%
We may choose%
\begin{equation}
\tau ^{\prime }=-2\sqrt{a}<0\qquad \text{and\qquad }\lambda =b-2\sqrt{ac}-%
\sqrt{a} . \label{STP4}
\end{equation}

As a result, for all Sommerfeld-type potentials, by the Nikiforov-Uvarov
quantization rule (\ref{NU7}) one obtains%
\begin{equation}
\dfrac{b}{2\sqrt{a}}-\sqrt{c}=n+\dfrac{1}{2},  \label{STP5}
\end{equation}%
as an equation for the exact energy levels. (It is worth noting that
Sommerfeld had obtained a similar relation in special cases \cite{SomQM}.)

\subsection{The puzzle resolution}

By (\ref{WKB8}) and (\ref{STP5}), we arrive at the following result.

\begin{theorem}
\begin{equation}
a=A,\qquad b=B,\qquad c=C.  \label{Resolution}
\end{equation}
\end{theorem}

Thus, the equations for energy levels are identical in both, exact and
approximate, approaches for all central potentials discussed by Sommerfeld.
According to our analysis, the equivalence of the energy levels for all
Sommerfeld-type potentials occurs as a result of mathematical coincidence as
earlier stated by Schr\"{o}dinger \cite{Biedenharn}, \cite{YouMan} (see
below).

On the contrary, Heisenberg writes \cite{Heisenberg}: \textquotedblleft {%
\textit{It would be intriguing to explore whether this is about a miracle or
it is the group-theoretical approach which leads to this formula}}%
.\textquotedblright\

\section{Examples}

We discuss several typical cases when the energy levels coincide in both,
exact and approximate, approaches for the Sommerfeld-type potentials.

\subsection{Nonrelativistic Coulomb problem}

In the well-known case of nonrelativistic Coulomb's problem, one gets%
\begin{eqnarray}
&&u^{\prime \prime }+\left[ 2\left( \varepsilon _{0}+\frac{Z}{x}\right) -%
\frac{\left( l+1/2\right) ^{2}}{x^{2}}\right] u=0  \label{NRC1} \\
&&\left( \varepsilon _{0}=\dfrac{E}{E_{0}},\,\, E_{0}=\dfrac{e^{2}}{%
a_{0}},\,\, a_{0}=\frac{\hbar ^{2}}{me^{2}},\,\, x=\frac{r}{a_{0}}\right)  \notag
\end{eqnarray}%
(in dimensionless units with Langer's modification).

Thus, $A=-2\varepsilon _{0}\/,$ $B=2Z\/,$ $C=(l+1/2)^{2}\/$ and in view of
the quantization rule \cite{SomQM}:
\begin{equation}
\dfrac{Z}{\sqrt{-2\varepsilon _{0}}}-l-\dfrac{1}{2}=n_{r}+\dfrac{1}{2}\/.
\label{NRC2}
\end{equation}%
As a result, we obtain exact energy levels for the nonrelativistic
hydrogenlike problem:
\begin{equation}
\varepsilon _{0}=\dfrac{E}{E_{0}}=-\dfrac{Z^{2}}{2(n_{r}+l+1)^{2}}\/.
\label{NRC3}
\end{equation}%
Here, $n=n_{r}+l+1$ is the principal quantum number \cite{LaLif3}.

\subsection{Relativistic Schr\"{o}dinger equation}

In the case of the relativistic Schr\"{o}dinger equation, one should write
\cite{Barleyetal2021}, \cite{Dav}, \cite{NU}, \cite{Schiff}, \cite{SomQM}:
\begin{eqnarray}
&&u^{\prime \prime }+\left[ \left( \varepsilon +\frac{\mu }{x}\right) ^{2}-1-%
\frac{\left( l+1/2\right) ^{2}}{x^{2}}\right] u=0  \label{RCP1} \\
&&\qquad \left( \varepsilon =\frac{E}{mc^{2}},\quad \mu =\frac{Ze^{2}}{\hbar
c},\quad x=\frac{mc}{\hbar }r\right) .  \notag
\end{eqnarray}%
Here, $A=1-{\varepsilon }^{2}\/,$ $B=2\mu \varepsilon \/,$ and $%
C=(l+1/2)^{2}-{\mu }^{2}\/$ (with Langer's modification). The combined
quantization rule implies
\begin{equation}
\dfrac{\mu \varepsilon }{\sqrt{1-{\varepsilon }^{2}}}=n_{r}+\nu +1\/,\quad
\nu =-\frac{1}{2}+\sqrt{\left( l+\frac{1}{2}\right) ^{2}-\mu ^{2}}
\label{RCP2}
\end{equation}%
and the formula for the relativistic energy levels is given by \cite{Dav},
\cite{Schiff}, \cite{SomQM}:%
\begin{eqnarray}
\frac{E_{n_{r},\,l}}{mc^{2}} &=&\frac{1}{\sqrt{1+\dfrac{\mu ^{2}}{\left[
n_{r}+\frac{1}{2}+\sqrt{\left( l+\frac{1}{2}\right) ^{2}-\mu ^{2}}\right]
^{2}}}}  \label{RCP3} \\
&=&1-\frac{\mu ^{2}}{2n^{2}}-\frac{\mu ^{4}}{2n^{4}}\left( \frac{n}{l+1/2}-%
\frac{3}{4}\right) +\text{O}\left( \mu ^{6}\right) ,  \notag
\end{eqnarray}%
where the expansion, in the limit $\mu =(Ze^{2})/(\hbar c)\rightarrow 0,$
when $ c\rightarrow \infty,$ can be derived by a direct Taylor's formula and/or verified by a computer
algebra system. Here, $n=n_{r}+l+1$ is the corresponding nonrelativistic
principal quantum number. The first term in this expansion is simply the
rest mass energy $E_{0}=mc^{2}$ of the charged spin-zero particle, the
second term coincides with the energy eigenvalue in the nonrelativistic Schr%
\"{o}dinger theory and the third term gives the so-called fine structure of
the energy levels, which removes the degeneracy between states of the same $%
n $ and different $l\/.$

\subsection{Dirac equation}

Sommerfeld's fine structure formula for the relativistic Coulomb problem
\cite{Som1916}, \cite{SomAS}, can be thought of as the main achievement of
the `old' quantum mechanics. Here, we will derive this result in a
semiclassical approximation for the radial Dirac equations (separation of
variables in spherical coordinates is discussed in detail elsewhere; see for
instance \cite{AkhBer}, \cite{BerLifPit}, \cite{Darwin}, \cite{Gordon}, \cite%
{Susetal20}). In the dimensionless units, one of these second-order
differential equations has the form
\begin{equation}
v_{1}^{\prime \prime }+\dfrac{({\varepsilon }^{2}-1)x^{2}+2\varepsilon \mu
x-\nu (\nu +1)}{x^{2}}\,v_{1}=0\/  \label{Dirac1}
\end{equation}%
and the second equation can be obtained from the first one by replacing $\nu
\rightarrow -\nu \/$. By Langer's modification, we obtain
\begin{equation}
p(x)=\left[ \left( \varepsilon +\frac{\mu }{x}\right) ^{2}-1-\frac{\left(
\nu +1/2\right) ^{2}+\mu ^{2}}{x^{2}}\right] ^{1/2}\/.  \label{Dirac2}
\end{equation}%
For the Dirac equation, $A=1-{\varepsilon }^{2}\/,$ $B=2\mu \varepsilon \/,$
$C=(\nu +1/2)^{2}$. The relativistic energy levels of an electron in the
central Coulomb field are given by \cite{Darwin}, \cite{Gordon}, \cite%
{Som1916}, \cite{SomAS}:
\begin{equation}
E=E_{n_{r},\,j}=\frac{mc^{2}}{\sqrt{1+\mu ^{2}/\left( n_{r}+\nu \right) ^{2}}%
}\quad (n_{r}=0,1,2,\ ...).  \label{Dirac3}
\end{equation}%
Here, $\mu =(Ze^{2})/(\hbar c)$ and in Dirac's theory,%
\begin{equation}
\nu =\nu _{\text{Dirac}}=\sqrt{\left( j+1/2\right) ^{2}-\mu ^{2}},
\label{Dirac4}
\end{equation}%
where $j=1/2,3/2,5/2,\ ...\ $ is the total angular momentum including the
spin of the relativistic electron.

In Dirac's theory the nonrelativistic limit has the form%
\begin{equation}
\frac{E_{n_{r},\,j}}{mc^{2}}=1-\frac{\mu ^{2}}{2n^{2}}-\frac{\mu ^{4}}{2n^{4}%
}\left( \frac{n}{j+1/2}-\frac{3}{4}\right) +\text{O}\left( \mu ^{6}\right)
,\quad \mu \rightarrow 0,  \label{Dirac5}
\end{equation}%
where $n=n_{r}+j+1/2$ is the principal quantum number of the nonrelativistic
hydrogenlike atom. Once again, the first term in this expansion is the rest
mass energy of the relativistic electron, the second term coincides with the
energy eigenvalue in the nonrelativistic Schr\"{o}dinger theory and the
third term gives the so-called fine structure of the energy levels --- the
correction obtained for the energy in the Pauli approximation which includes
the interaction of the spin of the electron with its orbital angular
momentum. The total spread in the energy of the fine-structure levels is in
agreement with experiments.

{The maximum spreads of the fine-structure levels occur when $l=0\/,$ $l=n-1$
and $j=1/2\/,$ $j=n-1/2$ for the }Schr\"{o}dinger (\ref{RCP3}) and Dirac (%
\ref{Dirac5}) theories{, respectively. Therefore, for the quotient, one gets%
\begin{equation}
\frac{\Delta E_{\text{Schr\"{o}dinger}}}{\Delta E_{\text{Sommerfeld}}}=\frac{%
4n}{2n-1}\qquad (n=2,3,\,...)\/.  \label{Dirac6}
\end{equation}%
} (See {\cite{Barleyetal2021}, \cite{Dav}, \cite{Schiff}, \cite{SomQM} for
more details.)}

In connection with Sommerfeld's fine-structure formula Erwin Schr\"{o}%
\-dinger writes, \textit{inter alia\/}, in a letter dated 29th February,
1956 \cite{Biedenharn}, \cite{YouMan}: \textit{{\ {\textquotedblleft ... you
are naturally aware of the fact that Sommerfeld derivation of the
fine-structure formula provides only fortuitously the result demanded by the
experiment. One may notice then from this particular example that newer form
of quantum theory (i.e., quantum mechanics) is by no means such an
inevitable continuation of the older theory as is commonly supposed.
Admittedly the Schr\"{o}dinger theory, relativistically framed (without
spin), gives a {formal} expression of the fine-structure formula of
Sommerfeld, but it is {incorrect} owing to the appearance of half-integers
instead of integers. My paper in which this is shown has ... never been
published; it was withdrawn by me and replaced by non-relativistic
treatment... The computation [by the relativistic method] is far too little
known. It shows in one respect how {necessary} Dirac's improvement was, and
on the other hand it is wrong to assume that the older form of quantum
theory is `broadly' in accordance with the newer form. \/\textquotedblright }%
}}

It took two quantum revolutions, from 1916 until 1928, in order to derive
Sommerfeld's formula in the Dirac theory of the relativistic hydrogen atom!

\subsection{Kratzer potential}

In order to investigate the vibrational-rotational spectrum of a diatomic
molecule, the following potential%
\begin{equation}
U(r)=-2D\left( \dfrac{a}{r}-\dfrac{1}{2}\dfrac{a^{2}}{r^{2}}\right) \/ , \qquad
D>0\/,  \label{KMP1}
\end{equation}%
with a minimum $U(a)=-D\/$ and $0<r<\infty ,$ has been used \cite{EEKS},
\cite{Flugge}.

We are looking for solutions of the Schr\"{o}dinger equation in spherical
coordinates and introduce the dimensionless quantities:%
\begin{equation}
x=\dfrac{r}{a},\qquad \beta ^{2}=-\dfrac{2ma^{2}}{\hbar ^{2}}\ E,\qquad
\gamma ^{2}=\dfrac{2ma^{2}}{\hbar ^{2}}\ D  \label{KMP2}
\end{equation}%
together with the standard substitution: $R(r)=u(x)\/.$

For bound states $E<0\/,\beta >0\/$ and the radial equation takes the form%
\begin{equation}
u^{\prime \prime }+\left[ -\beta ^{2}+\frac{2\gamma ^{2}}{x}-\frac{\gamma
^{2}+l(l+1)}{x^{2}}\right] u=0\/.  \label{KMP3}
\end{equation}%
Here, $A=-\beta ^{2}\/$, $B=2\gamma ^{2}\/$, and $C=\gamma ^{2}+\left(
l+1/2\right) ^{2}\/$ with Langer's modification.

As a result, the bound states are given by%
\begin{equation}
E_{n,l}=-\dfrac{2ma^{2}D^{2}}{\hbar ^{2}}\,\dfrac{1}{(\nu +n)^{2}}\/,
\label{KMP4}
\end{equation}%
where%
\begin{equation}
\nu =\dfrac{1}{2}+\sqrt{\gamma ^{2}+\left( l+\dfrac{1}{2}\right) ^{2}}%
,\qquad \gamma ^{2}=\dfrac{2ma^{2}}{\hbar ^{2}}\ D\/.  \label{KMP5}
\end{equation}%
(For further details and applications, see \cite{EEKS}, \cite{Flugge}, and
the references therein.)

\section{$n$-Dimensional Problems}

\subsection{Separation of variables}

For hyperspherical coordinates in $%
\mathbb{R}
^{n},$ when $\mathbf{x=}r\mathbf{s},\ \mathbf{s}^{2}=1,$ the Laplace
operator take the form \cite{NSU}:%
\begin{equation}
\Delta =\Delta _{r}+\frac{1}{r^{2}}\Delta _{\mathbf{s}},  \label{nDSep1}
\end{equation}%
where%
\begin{equation}
\Delta _{r}=\frac{1}{r^{n-1}}\frac{\partial }{\partial r}\left( r^{n-1}\frac{%
\partial }{\partial r}\right)  \label{nDSep2}
\end{equation}%
and%
\begin{equation}
\Delta _{\mathbf{s}}Y+\lambda Y=0,\quad \lambda =l\left( l+n-2\right) ,\quad
l=0,1,2,~\ldots  \label{nDSep3}
\end{equation}%
for a set of hyperspherical harmonics $Y\left( \mathbf{s}\right)
=Y_{l,\left\{ l_{k}\right\} }\left( \mathbf{s}\right) $ corresponding to a
given tree \cite{NSU}, \cite{Varshalovich1988}.

The stationary Schr\"{o}dinger equation, in a central field,%
\begin{equation}
\widehat{H}\Psi =E\Psi ,\quad \widehat{H}=-\frac{\hbar ^{2}}{2m}\Delta
+V\left( r\right) ,  \label{nDSep4}
\end{equation}%
admits a separation of the variables in hyperspherical coordinates:%
\begin{equation}
\Psi =R\left( r\right) Y\left( \mathbf{s}\right) .  \label{nDSep5}
\end{equation}

The radial equation in $%
\mathbb{R}
^{n}$ when $\mathbf{x=}r\mathbf{s},\mathbf{s}^{2}=1,$ take the form%
\begin{eqnarray}
&&\frac{1}{r^{n-1}}\frac{d}{dr}\left( r^{n-1}\frac{dR}{dr}\right) -\frac{%
l(l+n-2)}{r^{2}}R  \label{nDSep6} \\
&&\quad \qquad \qquad \qquad +\frac{2m}{\hbar ^{2}}\left[ E-V(r)\right] R=0 \/ .
\notag
\end{eqnarray}

The following identities%
\begin{eqnarray}
\frac{1}{r^{n-1}}\frac{d}{dr}\left( r^{n-1}\frac{dR}{dr}\right)  &=&\frac{1}{%
r^{(n-1)2}}\left( r^{(n-1)/2}R\right) ^{\prime \prime }  \label{nDSep7} \\
&&-\frac{(n-1)(n-3)}{4r^{2}}R \/ \notag
\end{eqnarray}

and%
\begin{eqnarray}
l(l+n-2)+\frac{(n-1)(n-3)}{4} &=&\left( l+\frac{n-3}{2}\right) \left( l+%
\frac{n-1}{2}\right)  \label{nDSep8} \\
&=&\left( l+\frac{n-2}{2}\right) ^{2}-\frac{1}{4}  \notag
\end{eqnarray}%
are easily verified. As a result, if $\chi =$ $r^{(n-1)/2}R:$%
\begin{equation}
\chi ^{\prime \prime }+\left[ \frac{2m}{\hbar ^{2}}\left( E-V(r)\right) -%
\frac{\left( l+(n-3)/2\right) (l+(n-1)/2)}{r^{2}}\right] \chi =0
\label{nDSep9}
\end{equation}%
as an extension of the radial equation in $%
\mathbb{R}
^{3}$ to $%
\mathbb{R}
^{n}.$ The following special cases of the centrifugal term were discussed by
Sommerfeld \cite{SomQM}:%
\begin{equation}
\left( l+\frac{n-3}{2}\right) \left( l+\frac{n-1}{2}\right) =\left\{
\begin{array}{cc}
0,\ n=1,l=0 & \text{in }%
\mathbb{R}
^{1} \\
l^{2}-\frac{1}{4},\ n=2 & \text{in }%
\mathbb{R}
^{2} \\
l(l+1),\ n=3 & \text{in }%
\mathbb{R}
^{3}%
\end{array}%
\right. .  \label{nDSep10}
\end{equation}%
One can see that the radial equation (\ref{nDSep9}) can be solved for the
Sommer\-feld-type potentials, thus extending the puzzle to $%
\mathbb{R}
^{n}.$ Moreover, in the radial equation, $%
\mathbb{R}
^{3}\rightarrow
\mathbb{R}
^{n}$ provided%
\begin{equation}
l\rightarrow l+\frac{n-3}{2},  \label{nDSep11}
\end{equation}%
which allows to extend all exactly solvable in $%
\mathbb{R}
^{3}$ potentials into $%
\mathbb{R}
^{n}.$

\subsection{Kepler problems}

For the nonrelativistic Coulomb problem in $%
\mathbb{R}
^{n}$, one gets in dimensionless units:%
\begin{eqnarray}
&&u^{\prime \prime }+\left[ 2\left( \varepsilon _{0}+\frac{Z}{x}\right) -%
\frac{\left( l+(n-3)/2\right) \left( l+(n-1)/2\right) }{x^{2}}\right] u=0
\notag \\
&&\quad \left( \varepsilon _{0}=\dfrac{E}{E_{0}},\,\,\,E_{0}=\dfrac{e^{2}}{%
a_{0}},\,\,\,a_{0}=\frac{\hbar ^{2}}{me^{2}}\right) .  \label{nDKep1}
\end{eqnarray}

With the Langer modification: $A=-2\varepsilon _{0}\/,$ $B=2Z\/,$ $%
C=(l+(n-2)/2)^{2}\/$ and in view of the quantization rule (\ref{WKB8}) one
gets:%
\begin{equation}
\dfrac{Z}{\sqrt{-2\varepsilon _{0}}}-l-\dfrac{n-2}{2}=n_{r}+\dfrac{1}{2}\/.
\label{nDKep2}
\end{equation}%
As a result, we obtain exact energy levels for the $n$-dimensional
nonrelativistic hydrogenlike problem:%
\begin{equation}
\varepsilon _{0}=\dfrac{E}{E_{0}}=-\dfrac{Z^{2}}{2(n_{r}+l+(n-1)/2)^{2}}\/
\label{nDKep3}
\end{equation}%
in the WKB approximation, which is identical to the exact solution.

\subsection{Harmonic oscillators}

In $%
\mathbb{R}
^{n},$ when $\mathbf{x=}r\mathbf{s},\ \mathbf{s}^{2}=1,$ and%
\begin{equation}
V(r)=\frac{1}{2}m\omega ^{2}r^{2}  \label{nDHar1}
\end{equation}%
the radial equation takes the form%
\begin{equation}
\chi ^{\prime \prime }+\left[ \frac{2m}{\hbar ^{2}}\left( E-\frac{1}{2}%
m\omega ^{2}r^{2}\right) -\frac{\left( l+(n-3)/2\right) (l+(n-1)/2)}{r^{2}}%
\right] \chi =0 \/ . \label{nDHar2}
\end{equation}%
The Bohr-Sommerfeld quantization rule, with Langer's modification, is given
by%
\begin{eqnarray}
&&\int_{r_{1}}^{r_{2}}\sqrt{\frac{2m}{\hbar ^{2}}\left( E-\frac{1}{2}m\omega
^{2}r^{2}\right) -\frac{\left( l+(n-2)/2\right) ^{2}}{r^{2}}}\ dr
\label{nDHar3} \\
&&\qquad \qquad \qquad =\pi \left( n_{r}+\frac{1}{2}\right) \/ . \notag
\end{eqnarray}

Transforming the integral into a Sommerfeld-type form,%
\begin{equation}
\int_{r_{1}}^{r_{2}}\sqrt{\frac{2mE}{\hbar ^{2}r^{2}}-\frac{m^{2}\omega ^{2}%
}{\hbar ^{2}}-\frac{\left( l+(n-2)/2\right) ^{2}}{r^{4}}}\ \left( rdr\right)
,  \label{nDHar4}
\end{equation}%
one can introduce the new variable $\xi =r^{2}$ and conclude that%
\begin{equation}
A=\frac{m^{2}\omega ^{2}}{4\hbar ^{2}},\quad B=\frac{mE}{2\hbar ^{2}},\quad
C=\frac{1}{4}\left( l+\frac{n-2}{2}\right) ^{2}.  \label{nDHar5}
\end{equation}%
As a result, in WKB approximation, the energy levels are given by%
\begin{equation}
E_{n_{r}}=\hbar \omega \left( 2n_{r}+l+\frac{1}{2}\right) ,\quad
n_{r}=0,1,2,\ldots \ .  \label{nDHar6}
\end{equation}

On the contrary, introducing the dimensionless units%
\begin{equation}
\frac{\kappa ^{2}}{2\mu }=\varepsilon =\frac{E}{\hbar \omega },\quad \mu =%
\frac{m\omega }{\hbar },\quad \xi =\mu r^{2}  \label{nDHar7}
\end{equation}%
we arrive at the generalized equation of hypergeometric type with the
following coefficients%
\begin{eqnarray}
\sigma (\xi ) &=&\xi ,\qquad \widetilde{\tau }(\xi )=\frac{1}{2},
\label{nDHar8} \\
\widetilde{\sigma }(\xi ) &=&\frac{1}{4}\left[ (2\varepsilon )\xi -\xi
^{2}-\left( l+\frac{n-3}{2}\right) \left( l+\frac{n-1}{2}\right) \right]
\notag
\end{eqnarray}%
and the Nikiforov-Uvarov approach results in the same eigenvalues.

Indeed, one may choose, in a generic form,%
\begin{eqnarray}
\sigma (\xi ) &=&\xi ,\qquad \widetilde{\tau }=\frac{1}{2}  \label{nDHar9} \\
\widetilde{\sigma }(\xi ) &=&b\xi -a\xi ^{2}-c+\frac{1}{16}.  \notag
\end{eqnarray}%
Then%
\begin{equation}
\pi (\xi )=\frac{1}{2}\pm \left( \sqrt{a}~\xi \pm \sqrt{c}\right) ,\quad
k-b\pm 2\sqrt{ac}.  \label{nDHar10}
\end{equation}%
We may choose $k=b-2\sqrt{ac}$ and $\pi (\xi )=\frac{1}{2}+\sqrt{c}-\sqrt{a}%
\ \xi .$ Then%
\begin{eqnarray}
\lambda &=&k+\pi ^{\prime }=b-2\sqrt{ac}-\sqrt{a}  \label{nDHar11} \\
\tau (\xi ) &=&\widetilde{\tau }+2\pi =\frac{3}{2}+2\sqrt{c}-2\sqrt{a}\ \xi
\notag
\end{eqnarray}%
and the quantization rule, once again, is given by%
\begin{equation}
\frac{b}{2\sqrt{a}}-\sqrt{c}=n_{r}+\frac{1}{2} \/ . \label{nDHar12}
\end{equation}

For the WKB solutions, in the radial equation:%
\begin{equation}
u^{\prime \prime }+\frac{1}{2\xi }u^{\prime }+\frac{b\xi -a\xi ^{2}-c+1/16}{%
\xi ^{2}}u=0,  \label{nDHar13}
\end{equation}
in the dimensionless units, one can use a back substitution $u\left( \xi
\right) =v\left( \eta \right) $ and $\eta =\sqrt{\xi }.$ Then%
\begin{equation}
v^{\prime \prime }+4\frac{b\eta ^{2}-a\eta ^{4}-c+1/16}{\eta ^{2}}v=0.
\label{nDHar14}
\end{equation}%
With the Langer-type modification, the Bohr-Sommerfeld rule reads%
\begin{eqnarray}
\pi \left( n_{r}+\frac{1}{2}\right) &=&\int_{\eta _{1}}^{\eta _{2}}p\left(
\eta \right) \ d\eta =2\int_{\eta _{1}}^{\eta _{2}}\sqrt{b-a\eta ^{2}-\frac{c%
}{\eta ^{2}}}\ d\eta  \label{nDHar15} \\
&=&\int_{\eta _{1}}^{\eta _{2}}\sqrt{-a+\frac{b}{\eta ^{2}}-\frac{c}{\eta
^{4}}}\ 2(\eta d\eta )  \notag \\
&=&\int \sqrt{-a+\frac{b}{\xi }-\frac{c}{\xi ^{2}}}\ d\xi =\pi \left( \frac{b%
}{2\sqrt{a}}-\sqrt{c}\right)  \notag
\end{eqnarray}%
and we arrive to the energy equation (\ref{nDHar12}) once again. Use%
\begin{equation}
a=\frac{1}{4},\quad b=\frac{\varepsilon }{2},\quad c=\frac{1}{4}\left( l+%
\frac{n-2}{2}\right) ^{2}  \label{nDHar16}
\end{equation}%
in order to obtain the corresponding exact energy levels (\ref{nDHar6}).

\section{Some Extensions}

\subsection{Trigonometric case: P\"{o}schl-Teller potential hole}

The following generic integral,%
\begin{equation}
I\left( A\right) =\int_{\theta _{1}}^{\theta _{2}}\sqrt{A-\frac{B}{\cos
^{2}\theta }-\frac{C}{\sin ^{2}\theta }}\ d\theta ,  \label{PTP1}
\end{equation}%
with the aid of substitution $T=\cos 2\theta ,$ can be transformed into the
sum of two similar Sommerfeld-type integrals:%
\begin{eqnarray}
I\left( A\right)  &=&\frac{\pi }{2\sqrt{2}}\left( \frac{\left( A-B+C\right)
}{2\sqrt{A}}-\sqrt{2B}\right)   \notag \\
&&+\frac{\pi }{2\sqrt{2}}\left( \frac{\left( B-C+A\right) }{2\sqrt{A}}-\sqrt{%
2C}\right)   \notag \\
&=&\frac{\pi }{2}\left( \sqrt{A}-\sqrt{B}-\sqrt{C}\right) .  \label{PTP2}
\end{eqnarray}%
(For an independent evaluation, see also appendix below.) As a result, the
Bohr-Sommerfeld quantization rule gives%
\begin{equation}
\sqrt{A}=\sqrt{B}+\sqrt{C}+2n+1.  \label{PTP3}
\end{equation}

The stationary Schr\"{o}dinger equation for the familiar P\"{o}schl-Teller
potential takes the form \cite{EEKS}, \cite{Flugge}:%
\begin{equation}
\frac{d^{2}\psi }{dx^{2}}+\frac{2m}{\hbar ^{2}}\left[ E-U\left( x\right) %
\right] \psi =0,  \label{PTP4}
\end{equation}%
where $0< x < \pi/(2\alpha)$ and%
\begin{equation}
U\left( x\right) =\frac{V_{0}}{2}\left[ \frac{a\left( a-1\right) }{\sin
^{2}\left( \alpha x\right) }+\frac{b\left( b-1\right) }{\cos ^{2}\left(
\alpha x\right) }\right] ,\quad V_{0}=\frac{\hbar ^{2}\alpha ^{2}}{m}.
\label{PTP5}
\end{equation}%
Here, with the help of the Langer-type modification: $a\left( a-1\right)
\rightarrow \left( a-1/2\right) ^{2}$ and $b\left( b-1\right) \rightarrow
\left( b-1/2\right) ^{2},$ one gets%
\begin{equation}
A=\frac{2mE}{\alpha ^{2}\hbar ^{2}},\quad B=\left( b-\frac{1}{2}\right)
^{2},\quad C=\left( a-\frac{1}{2}\right) ^{2}  \label{PTP6}
\end{equation}%
and\noindent
\begin{equation}
E_{n}=\frac{V_{0}}{2}\left( a+b+2n\right) ^{2}.  \label{PTP7}
\end{equation}%
In our consideration, the WKB method results in the exact energy levels
derived in \cite{EEKS}, \cite{Flugge}. (The mathematical motivation of the
Langer-type modification will be discussed elsewhere.)

\noindent \textbf{Conclusion.\/} We have demonstrated that for a large class
of exactly solvable potentials the approximate WKB energy levels coincide
with exact ones. Most of these problems have different symmetry groups.

\subsection{Hyperbolic case: modified P\"{o}schl-Teller potential hole}

\noindent For a modified potential,%
\begin{equation}
U\left( x\right) =-\frac{V_{0}}{\cosh ^{2}\left( \alpha x\right) } \qquad (-\infty < x < \infty),
\label{MPT1}
\end{equation}%
the Bohr-Sommerfeld quantization rule with the help of the integral (\ref%
{HI3}) below result in $(\varepsilon =-E/V_{0}):$%
\begin{equation}
E_{n}=-\frac{\hbar ^{2}\alpha ^{2}}{2m}\left[ \frac{\sqrt{2mV_{0}}}{\hbar
\alpha }-\left( n+\frac{1}{2}\right) \right] ^{2},  \label{MPT2}
\end{equation}%
which is a bit different from the exact values given by
\begin{equation}
E_{n}=-\frac{\hbar ^{2}\alpha ^{2}}{2m}\left[ \frac{1}{2}\sqrt{\frac{8mV_{0}%
}{\hbar ^{2}\alpha ^{2}}+1}-\left( n+\frac{1}{2}\right) \right] ^{2}.
\label{MPT3}
\end{equation}%
(See \cite{EEKS}, \cite{Flugge}, \cite{Ghatetal}, for details.)

\section{Appendix: Integral Evaluations}

On a contrary, one can use a technique of differentiation with respect to
parameters for the familiar integrals related to the Bohr-Sommerfeld
quantization rule. As is well-known, if%
\begin{equation}
J\left( x\right) =\int_{f\left( x\right) }^{g\left( x\right) }F\left(
x,y\right) \ dy,  \label{J1}
\end{equation}%
then%
\begin{equation}
\frac{dJ}{dx}=\int_{f\left( x\right) }^{g\left( x\right) }\frac{\partial
F\left( x,y\right) }{\partial x}\ dy+F\left( x,g\left( x\right) \right)
\frac{dg}{dx}-F\left( x,f\left( x\right) \right) \frac{df}{dx}.  \label{J2}
\end{equation}%
In the WKB case, the last two terms vanish because the limits are the
turning points when the integrand equals zero \cite{Ghatetal}. In the
following examples, we utilize this procedure for the integrals occurring in
the \textquotedblleft Sommerfeld-type puzzle\/\textquotedblright\ cases
discussed in this note.
\medskip

\textbf{Example~1.\/} For the Sommerfeld-type integrals,%
\begin{equation}
I=\int_{r_{1}}^{r_{2}}p(r)\ dr,\qquad p(r)=\sqrt{-A+\dfrac{B}{r}-\dfrac{C}{%
r^{2}}}\quad \left( A\/,C>0\/\right) ,  \label{Som1}
\end{equation}%
provided $p(r_{1})=p(r_{2})=0\/,$ one gets%
\begin{eqnarray}
\frac{dI}{dB} &=&\frac{1}{2}\int_{r_{1}}^{r_{2}}\frac{dr}{\sqrt{-Ar^{2}+Br-C}%
}  \label{Som2} \\
&=&\frac{1}{2\sqrt{A}}\int_{r_{1}}^{r_{2}}\frac{dr}{\sqrt{\dfrac{B^{2}-4AC}{%
4A^{2}}-\left( r-\dfrac{B}{2A}\right) ^{2}}}  \notag \\
&=&\frac{1}{2\sqrt{A}}\left. \arcsin \left( \frac{2Ar-B}{\sqrt{B^{2}-4AC}}%
\right) \right\vert _{r_{1}}^{r_{2}}=\frac{\pi }{2\sqrt{A}}.  \notag
\end{eqnarray}%
As a result,%
\begin{equation}
\frac{dI}{dB}=\frac{\pi }{2\sqrt{A}},\qquad I\left( B_{0}=2\sqrt{AC}\right)
=0  \label{Som3}
\end{equation}%
and, by integration,%
\begin{equation}
I=\pi \left( \dfrac{B}{2\sqrt{A}}-\sqrt{C}\right) .\qquad \blacksquare
\label{Som4}
\end{equation}
\medskip

\textbf{Example~2.\/} Our trigonometric integral,%
\begin{eqnarray}
I\left( A\right)  &=&\int_{\theta _{1}}^{\theta _{2}}\sqrt{A-\frac{B}{\cos
^{2}\theta }-\frac{C}{\sin ^{2}\theta }}\ d\theta   \label{TI1} \\
&=&\frac{\pi }{2}\left( \sqrt{A}-\sqrt{B}-\sqrt{C}\right) ,  \notag
\end{eqnarray}%
can be evaluated in a similar fashion:%
\begin{equation}
\frac{dI}{dA}=\frac{\pi }{4\sqrt{A}},\quad I\left( A_{0}=\sqrt{B}+\sqrt{C}%
\right) =0.  \label{TI2}
\end{equation}%
Indeed, if $T=\cos 2\theta $ one gets%
\begin{equation}
I\left( A\right) :=\frac{1}{2}\int_{T_{1}}^{T_{2}}\sqrt{\left(
A-2B-2C\right) +2\left( B-C\right) T-AT^{2}} \, \frac{dT}{1-T^{2}} \/ , \label{TI3}
\end{equation}%
where%
\begin{eqnarray}
T_{1} &=&\frac{B-C-\sqrt{\left( A-B-C\right) ^{2}-4BC}}{A},  \label{TI4} \\
T_{2} &=&\frac{B-C+\sqrt{\left( A-B-C\right) ^{2}-4BC}}{A}.  \notag
\end{eqnarray}%
Thus%
\begin{eqnarray}
\frac{dI}{dA} &=&\frac{1}{4}\int_{T_{1}}^{T_{2}}\frac{dT}{\sqrt{\left(
A-2B-2C\right) +2\left( B-C\right) T-AT^{2}}}  \label{TI5} \\
&=&\frac{1}{4\sqrt{A}}\left. \arcsin \left( \frac{AT-B+C}{\sqrt{\left(
A-B-C\right) ^{2}-4BC}}\right) \right\vert _{T_{1}}^{T_{2}}=\frac{\pi }{4%
\sqrt{A}}.  \notag
\end{eqnarray}%
Simple integration results in%
\begin{equation}
I\left( A\right) =\frac{\pi }{2}\left( \sqrt{A}-\sqrt{B}-\sqrt{C}\right) ,
\label{TI6}
\end{equation}%
where we have used the initial condition (\ref{TI2}).$\quad \blacksquare $
\medskip

\textbf{Example~3.\/} The hypergeometric integral%
\begin{equation}
J\left( \varepsilon \right) =\int_{-x_{0}\left( \varepsilon \right)
}^{x_{0}\left( \varepsilon \right) }\sqrt{\sec \text{h}^{2}x-\varepsilon }%
~dx,\quad J\left( 1\right) =0 \/ . \label{HI1}
\end{equation}%
Here $0\leq \varepsilon \leq 1$ and $x_{0}\left( \varepsilon \right) =\sec $h%
$^{-1}\sqrt{\varepsilon }:$%
\begin{eqnarray}
\frac{dJ}{d\varepsilon } &=&-\frac{1}{2}\int_{-x_{0}\left( \varepsilon
\right) }^{x_{0}\left( \varepsilon \right) }\frac{dx}{\sqrt{\sec \text{h}%
^{2}x-\varepsilon }}  \label{HI2} \\
&=&-\frac{1}{2\sqrt{\varepsilon }}\left. \arcsin \left( \sqrt{\frac{%
\varepsilon }{1-\varepsilon }}\sinh x\right) \right\vert _{-x_{0}\left(
\varepsilon \right) }^{-x_{0}\left( \varepsilon \right) }  \notag \\
&=&-\frac{\pi }{2\sqrt{\varepsilon }},\qquad \qquad J\left( \varepsilon
\right) =\pi \left( 1-\sqrt{\varepsilon }\right) .  \notag
\end{eqnarray}%
Thus%
\begin{equation}
\int_{-x_{0}}^{x_{0}}\sqrt{\sec \text{h}^{2}x-\varepsilon }~dx=\pi \left( 1-%
\sqrt{\varepsilon }\right) .\qquad \blacksquare  \label{HI3}
\end{equation}%
(See also \cite{Ghatetal}.)
\medskip

\noindent \textbf{Acknowledgments\/.}
I am grateful to the organizers for their hospitality
and would like to thank Sergey Kryuchkov and Eugene Stepanov for their help.

\end{document}